\documentclass[twocolumn,superscriptaddress]{revtex4-1}
\pdfoutput=1
\usepackage{color}
\usepackage{graphicx}
\usepackage{epstopdf}
\usepackage{amsmath}
\usepackage[colorlinks,linkcolor=blue,anchorcolor=blue,citecolor=blue,urlcolor=blue]{hyperref}

\newcommand{\bq}{{\bf q}}
\newcommand{\bk}{{\bf k}}
\newcommand{\ba}{{\bf a}}
\newcommand{\mT}{\mathcal{T}}

\begin{document}
\title{$t_{2g}$-orbital model on a honeycomb lattice: application to antiferromagnet SrRu$_2$O$_6$}
\author{Da Wang}
\affiliation{National Laboratory of Solid State Microstructures $\&$ School of Physics, Nanjing
University, Nanjing, 210093, China}
\author{Wan-Sheng Wang}
\affiliation{National Laboratory of Solid State Microstructures $\&$ School of Physics, Nanjing
University, Nanjing, 210093, China}
\author{Qiang-Hua Wang}
\affiliation{National Laboratory of Solid State Microstructures $\&$ School of Physics, Nanjing
University, Nanjing, 210093, China}
\affiliation{Collaborative Innovation Center of Advanced Microstructures, Nanjing University, Nanjing 210093, China}
\begin{abstract}
Motivated by the recent discovery of high temperature antiferromagnet SrRu$_2$O$_6$ \cite{Hiley2014,Tian2015} and its potential to be the parent of a new superconductor upon doping,
we construct a minimal $t_{2g}$-orbital model on a honeycomb lattice to simulate its low energy band structure. Local Coulomb interaction is
taken into account through both random phase approximation and mean field theory.
Experimentally observed antiferromagnetic order is obtained in both approximations. In addition, our theory predicts that the magnetic moments on three $t_{2g}$-orbitals are
non-collinear as a result of the strong spin-orbit coupling of Ru atoms.
\end{abstract}
\maketitle
\section{Introduction}
Magnetism and superconductivity are closely related to each other, as a common thread in several families of unconventional superconductors. \cite{Scalapino2012} Singlet Cooper pairing is expected to be mediated by antiferromagnetic (AF) fluctuations near the AF phase boundary like in most cuprates \cite{Bednorz1986}, iron-based \cite{Kamihara2008}, and heavy fermion \cite{Steglich1979} superconductors. While triplet Cooper pairing is widely believed to be triggered by ferromagnetic fluctuations as in Sr$_2$RuO$_4$ \cite{Maeno1994,Huo2013,Wang2013}. Therefore, looking for unconventional superconductivity in materials with strong magnetic fluctuations is a guiding principle in the community. SrRu$_2$O$_6$ is synthesized recently \cite{Hiley2014} and reported as an antiferromagnet with a Neel temperature as high as $565$K \cite{Tian2015}. Similar to other ruthenate Sr$_{1+n}$Ru$_n$O$_{1+3n}$\cite{Grigera2001}, SrRu$_2$O$_6$ is also a layered material. Due to the layered property of the sample, the AF order may be easily destroyed by introducing quantum fluctuations via doping or high pressure. As a result, SrRu$_2$O$_6$ may be a good parent compound to realize high temperature superconductivity.

Different from ruthenate Sr$_{1+n}$Ru$_n$O$_{1+3n}$ in which the RuO$_6$ octahedra are point-sharing and form a square lattice \cite{Singh2001}, in SrRu$_2$O$_6$ the RuO$_6$ octahedra are edge-sharing and the Ru atoms are arranged on a honeycomb lattice. According to first-principle calculations, \cite{Tian2015,Singh2015} the $t_{2g}$-orbitals of Ru are found to dominate the low energy states. This suggests that a $t_{2g}$-orbital model on a honeycomb lattice is a relevant minimal model for further studies on the correlation effect. On the other hand, even though the spin and charge degrees of freedom have been broadly studied in the honeycomb lattice (such as graphene) with only $\pi$-electrons \cite{CastroNeto2009}, the orbital degrees of freedom would bring us new features. For instance, the studies of ($p_x,p_y$)-orbital models on the honeycomb lattice revealed Wigner crystallization \cite{Wu2007} and anomalous quantum Hall effect \cite{Wu2008,Zhang2014}. In parallel, SrRu$_2$O$_6$ provides us a natural realization of the $t_{2g}$ d-orbital on the honeycomb lattice.

In this paper, we first derive an effective $t_{2g}$-orbital tight-binding Hamiltonian as a minimal model for SrRu$_2$O$_6$. We then consider the correlation effect through both random phase approximation (RPA) and mean field theory. We obtain the experimentally observed AF order and estimate the Neel moment and transition temperature within the mean field theory. Furthermore, we find the orbital-resolved AF moments on three $t_{2g}$-orbitals are non-collinear as a result of the strong spin-orbit coupling (SOC) on Ru atoms. Our minimal model provides the basis for further theoretical studies, and the magnetic structure we uncovered would trigger further experimental interests in this new member of the ruthenate family.

\section{$t_{2g}$-orbital tight-binding hamiltonian}

\begin{figure}
\includegraphics[width=0.5\textwidth]{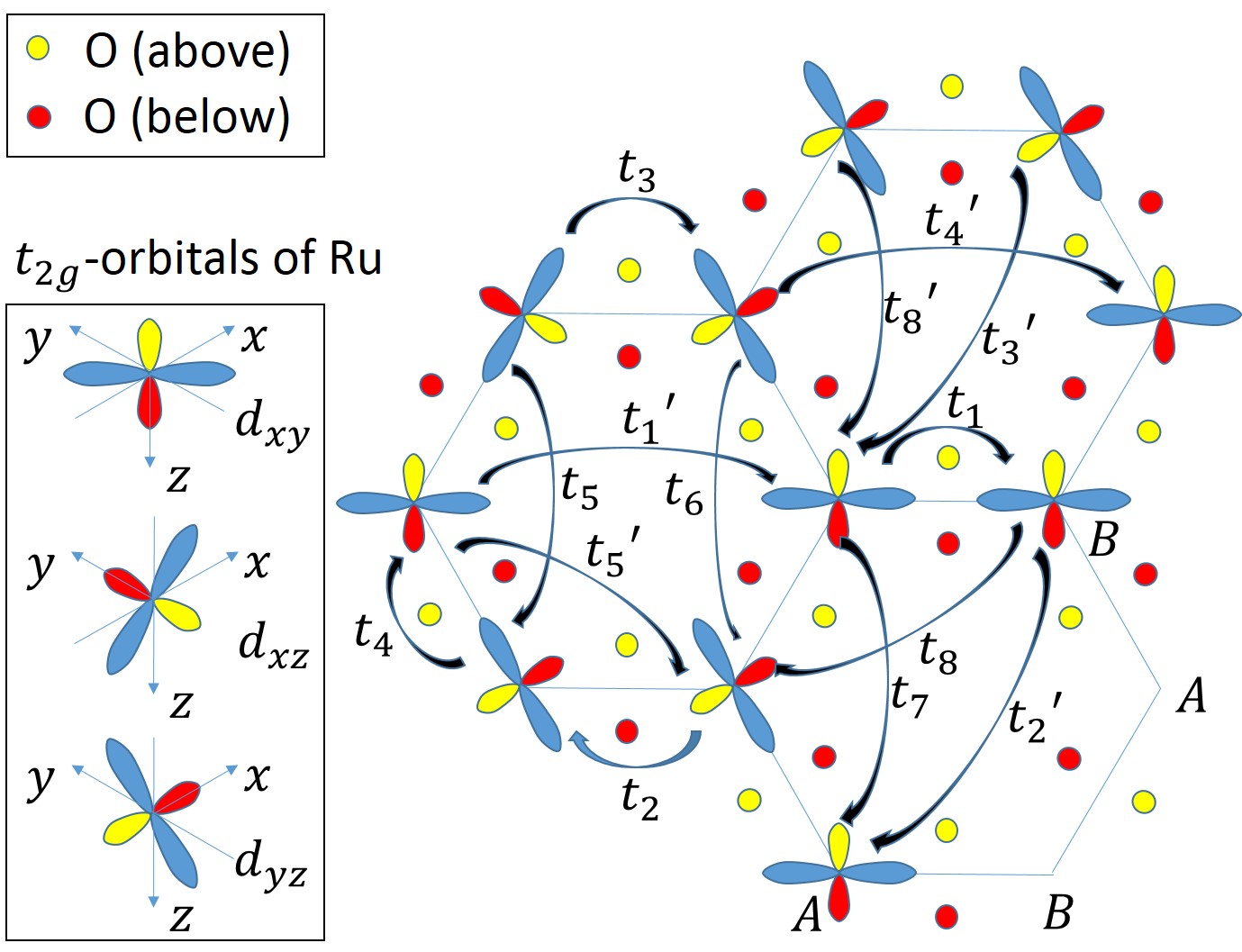}
\caption{Schematic discription of the lattice structure and all hopping elements up to the 3rd nearest neighbour. Three $t_{2g}$-orbitals of Ru are placed on the honeycomb lattice. O atoms (small circles) are distributed above (yellow) or below (red) the Ru-plane. Each RuO$_6$ octahedron is slightly distorted.}
\label{fig:hopping}
\end{figure}

In SrRu$_2$O$_6$, the low energy bands mainly come from the $t_{2g}$-orbitals of Ru atoms as reported by first-principle calculations. \cite{Tian2015,Singh2015} The Ru atoms form a honeycomb lattice in each RuO$_6$ layer. Therefore, to mimic the system by a minimal model, we only consider the $t_{2g}$-orbital electrons on a honeycomb lattice as shown in Fig.~\ref{fig:hopping}. The coordinate is set up with the origin at Ru-site and three axes point to three oxygen atoms above the Ru-plane in the undistorted RuO$_6$ octahedron. In this coordinate, the c-axis perpendicular to Ru-plane is along (1,1,1)-direction, and the $d_{xy}$, $d_{xz}$ and $d_{yz}$ orbitals forms an exact $t_{2g}$ multiplet, as schematically represented in Fig.~\ref{fig:hopping}. The blue/yellow/red lobes lie within/above/below the Ru-plane. Although in SrRu$_2$O$_6$, the RuO$_6$ octahedron is slightly twisted around and stretched along the c-axis, the $t_{2g}$-orbital degeneracy is protected by the $C_{3v}$ symmetry. As a result, the coordinate established in the undistorted octahedron will be used in this work.

To construct a tight-binding Hamiltonian, we keep nearest neighbor hopping $t_{1\sim4}$, next nearest neighbor hopping $t_{5\sim8,5',8'}$, and the third-neighbor hopping $t_{1'\sim4'}$. All these hopping elements are schematically shown in Fig.~\ref{fig:hopping}. We notice that $t_5\neq t_5'$ and $t_8\neq t_8'$ due to the distortion of the RuO$_6$ octahedra. Furthermore, we add intra- and inter-orbital on-site energies $V_0$ and $V_0'$, respectively. In a pure two dimensional model, $V_0'$ should be exactly zero due to the orthogonality of different orbitals. But here we are looking for an effective two dimensional model in the $k_c=0$ plane and thus the inter-layer hopping could lead to an effective on-site inter-orbital mixing $V_0'$. In addition, we add an SOC term $H_{\rm SOC}=-\lambda \sum_{i\mu} \psi_i^\dag L_\mu \otimes\sigma_\mu \psi_i$, where $\psi_i^t=[d_{i,xy,\uparrow},d_{i,xz,\uparrow},d_{i,yz,\uparrow},d_{i,xy,\downarrow},d_{i,xz,\downarrow},d_{i,yz,\downarrow}]$,  $\sigma_\mu=[\sigma_x,\sigma_y,\sigma_z]$ are three Pauli's matrices, and $L_\mu=[L_{x},L_{y},L_{z}]$ are rank-3 angular momentum matrices acting on orbital space, with all nonzero elements given by
\begin{eqnarray}
&& L_{x,12}=-L_{x,21}=-i, \quad L_{y,13}=-L_{y,31}=-i, \nonumber\\
&& L_{z,23}=-L_{z,32}=i.
\end{eqnarray}

With the above ingredients and for a given inter-layer momentum $k_c=0$,  we arrive at an effective two-dimensional tight-binding Hamiltonian $H=\sum_\bk \Psi_\bk^\dag H_{\bk} \Psi_\bk$ in the basis    $\Psi_\bk^t=[\psi_{A\bk\uparrow},\psi_{B\bk\uparrow},\psi_{A\bk\downarrow},\psi_{B\bk\downarrow}]$ where $\psi_{s,\bk\sigma}^t=[d_{s,xy,\bk\sigma},d_{s,xz,\bk\sigma},d_{s,yz,\bk\sigma}]$ for $s=A,B$ on the two sublattices. The matrix $H_{\bk}$ is explicitly written as

\begin{figure}
\includegraphics[width=0.45\textwidth]{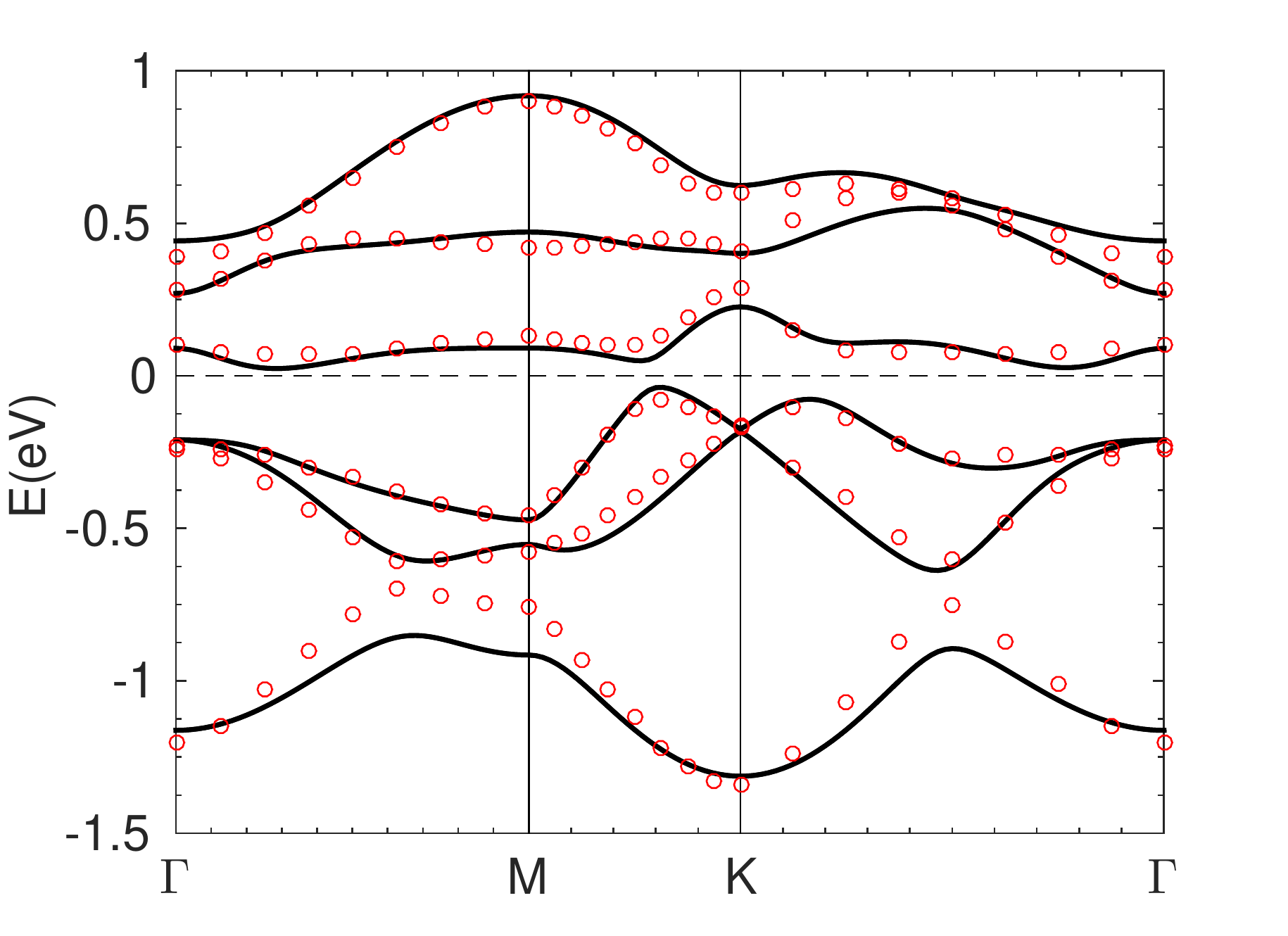}
\caption{Band dispersion of our $t_{2g}$-orbital tight-binding model along high symmetry lines. The local-density-approximation result (from Ref.~\onlinecite{Singh2015}) is also shown (red symbols) for comparison.}
\label{fig:band}
\end{figure}

\begin{widetext}
\begin{eqnarray}
H_\bk=\left[\begin{array}{cccc}
\mT_{AA}(\bk)+\mathcal{V}-\lambda L_z & \mT_{AB}(\bk) & -\lambda L_x+i\lambda L_y & 0 \\
\mT_{AB}(\bk)^\dag & \mT_{BB}(\bk)+\mathcal{V}-\lambda L_z & 0 & -\lambda L_x+i\lambda L_y \\
 -\lambda L_x-i\lambda L_y & 0 & \mT_{AA}(\bk)+\mathcal{V}+\lambda L_z & \mT_{AB}(\bk) \\
0 &  -\lambda L_x-i\lambda L_y & \mT_{AB}(\bk)^\dag & \mT_{BB}(\bk)+\mathcal{V}+\lambda L_z
\end{array}
\right] ,\nonumber\\
\label{eq:Hk}
\end{eqnarray}
in which
\begin{eqnarray}
T_{AB}(\bk)&=&\left[\begin{array}{ccc}t_1&t_4&t_4\\t_4&t_2&t_3\\t_4&t_3&t_2 \end{array}\right]{\rm e}^{i\bk\cdot \ba_1}
+ \left[\begin{array}{ccc}t_2&t_3&t_4\\t_3&t_2&t_4\\t_4&t_4&t_1 \end{array}\right]{\rm e}^{i\bk\cdot \ba_2}
+ \left[\begin{array}{ccc}t_2&t_4&t_3\\t_4&t_1&t_4\\t_3&t_4&t_2 \end{array}\right]{\rm e}^{i\bk\cdot \ba_3} ,\nonumber \\
&+&\left[\begin{array}{ccc}t_1'&t_4'&t_4'\\t_4'&t_2'&t_3'\\t_4'&t_3'&t_2' \end{array}\right]{\rm e}^{-2i\bk\cdot \ba_1}
+ \left[\begin{array}{ccc}t_2'&t_3'&t_4'\\t_3'&t_2'&t_4'\\t_4'&t_4'&t_1' \end{array}\right]{\rm e}^{-2i\bk\cdot \ba_2}
+ \left[\begin{array}{ccc}t_2'&t_4'&t_3'\\t_4'&t_1'&t_4'\\t_3'&t_4'&t_2' \end{array}\right]{\rm e}^{-2i\bk\cdot \ba_3} ,\nonumber\\
\mT_{AA}(\bk)&=&\left[\begin{array}{ccc}t_7&t_8&t_8'\\t_8'&t_6&t_5'\\t_8&t_5&t_6 \end{array}\right]{\rm e}^{i\bk\cdot (\ba_2-\ba_3)}
+ \left[\begin{array}{ccc}t_6&t_5'&t_8'\\t_5&t_6&t_8\\t_8&t_8'&t_7 \end{array}\right]{\rm e}^{i\bk\cdot (\ba_3-\ba_1)}
+ \left[\begin{array}{ccc}t_6&t_8&t_5\\t_8'&t_7&t_8\\t_5'&t_8'&t_6 \end{array}\right]{\rm e}^{i\bk\cdot (\ba_1-\ba_2)} +h.c.  ,\nonumber\\
\mT_{BB}(\bk)&=&\left[\begin{array}{ccc}t_7&t_8&t_8'\\t_8'&t_6&t_5'\\t_8&t_5&t_6 \end{array}\right]{\rm e}^{-i\bk\cdot (\ba_2-\ba_3)}
+ \left[\begin{array}{ccc}t_6&t_5'&t_8'\\t_5&t_6&t_8\\t_8&t_8'&t_7 \end{array}\right]{\rm e}^{-i\bk\cdot (\ba_3-\ba_1)}
+ \left[\begin{array}{ccc}t_6&t_8&t_5\\t_8'&t_7&t_8\\t_5'&t_8'&t_6 \end{array}\right]{\rm e}^{-i\bk\cdot (\ba_1-\ba_2)} +h.c. ,\nonumber\\
\end{eqnarray}
\end{widetext}
and
\begin{eqnarray}
\mathcal{V}=\left[\begin{array}{ccc}V_0&V_0'&V_0'\\V_0'&V_0&V_0'\\V_0'&V_0'&V_0\end{array}\right] ,
\end{eqnarray}
where $(\ba_1,\ba_2,\ba_3)$ are three displacements from an A-site to its nearest neighbour B-sites. All the model parameters are determined by fitting the first-principle band structure \cite{Singh2015}: (in unit of eV)
$t_1=0.16$, $t_2=-0.01$, $t_3=0.30$, $t_4=-0.02$,
$t_5=-0.10$, $t_5'=0$, $t_6=-0.01$, $t_7=0.04$, $t_8=0.11$, $t_8'=0.02$,
$t_1'=-0.04$, $t_2'=-0.01$, $t_3'=-0.01$, $t_4'=-0.01$, $V_0=-0.09$, $V_0'=-0.07$,
$\lambda=0.16$. In particular, we obtain the SOC strength $\lambda=0.16$eV, which is in agreement with the literature \cite{Mizokawa2001}. Based on these parameters, the band structure of our minimal model is plotted in Fig.~\ref{fig:band}.

In the paramagnetic state SrRu$_2$O$_6$ is a band insulator, as seen in our band structure and the first-principle result \cite{Singh2015,Tian2015}. However, a strong Hund's coupling would bind up the electrons to form a spin-$3/2$ state. Due to the bipartite lattice an AF order with moment $3\mu_B$/Ru is expected. Such an AF order has already been observed by Neutron scattering experiment in SrRu$_2$O$_6$,\cite{Tian2015}  but the observed moment is only $1.3\mu_B$/Ru, much smaller than $3\mu_B$. This indicates the inadequacy of a naive local moment picture. Instead, the itinerant property of electrons and SOC may play important roles. In the following, we will investigate the effect of correlation and SOC on the AF order in the itinerant picture of the $t_{2g}$-orbital model.

\section{Interaction and antiferromagnetism}

We adopt general multi-orbital local Coulomb interactions as follows,
\begin{eqnarray}
H_{\rm I}&=&\sum_{i}\left[\right. U\sum_{a} n_{ia\uparrow}n_{ia\downarrow}
+V\sum_{a > b} n_{ia}n_{ib} \nonumber\\
&+&J\sum_{a > b, \sigma\sigma'} a_{i\sigma}^\dag b_{i\sigma} b_{i\sigma'}^\dag a_{i\sigma'} +J'\sum_{a \neq b} a_{i\uparrow}^\dag a_{i\downarrow}^\dag
b_{i\downarrow}b_{i\uparrow} \left.\right] ,
\end{eqnarray}
where $n_{ia}=\sum_\sigma n_{ia\sigma}=\sum_\sigma a_{i\sigma}^\dag a_{i\sigma}$. $U$ is the Hubbard interaction, $V$ is the inter-orbital charge interaction, $J$ is the Hund's coupling and $J'$ is the pair hopping term.
These four interactions satisfies the relation $J'=J$ and $U=V+2J$. \cite{Castellani1978} Among these four terms, only $U$ and $J$ are responsible for magnetic channel instabilities. \cite{Raghu2008} So in the following discussions, we will only retain the $U$ and $J$ terms.

Since the non-interacting model is a band insulator, the bare susceptibility only depends on momentum very weakly. So we perform an RPA level calculation instead, since RPA will pick out relevant channels and strongly enhance their susceptibilities. 

\begin{figure}
\includegraphics[width=0.35\textwidth]{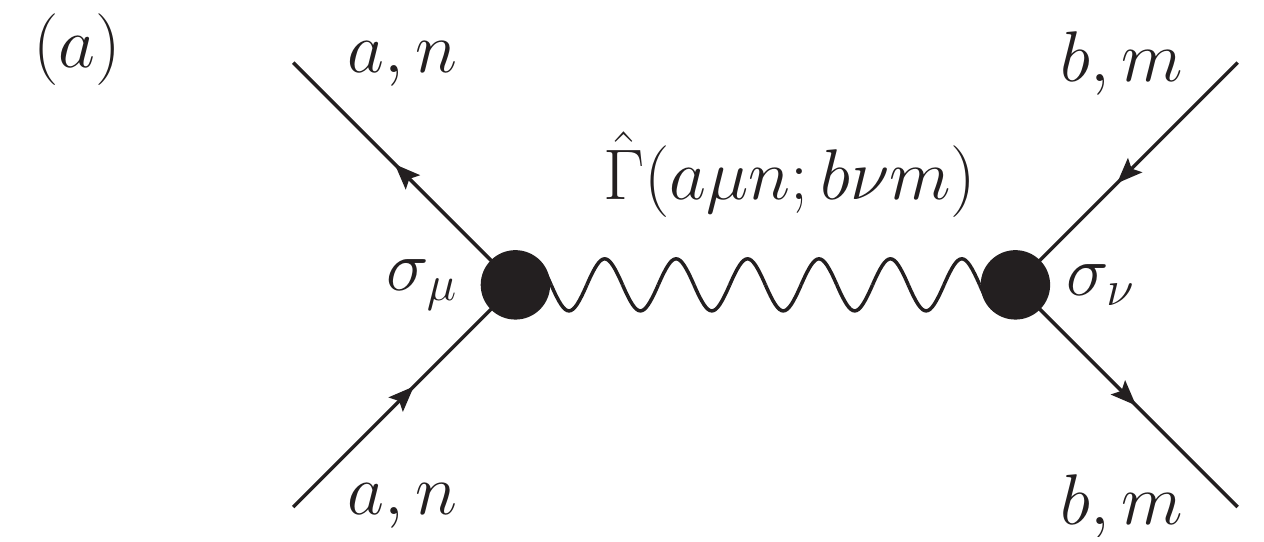}
\includegraphics[width=0.45\textwidth]{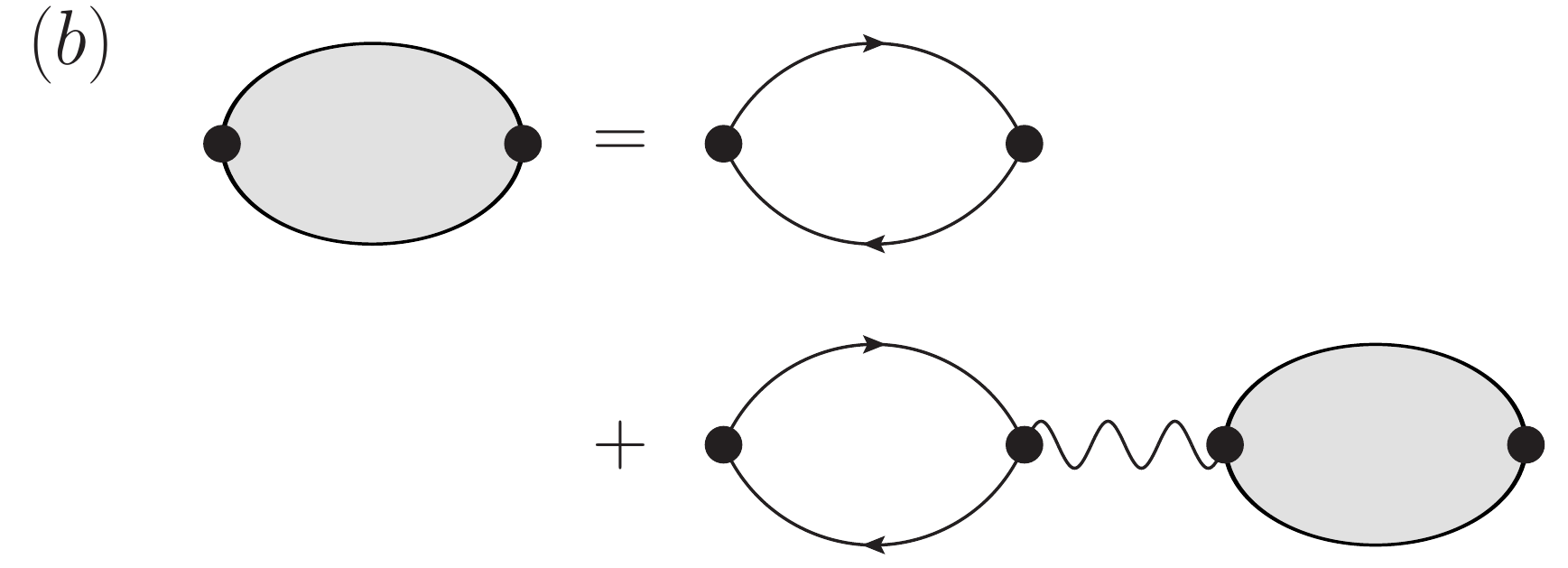}
\caption{(a)Vertex function $\hat{\Gamma}(a\mu n;b\nu m)$ defined in Eq.~\ref{eq:vertex} in the magnetic channel. (b)The Feynman diagrammatic representation of the magnetic susceptibility matrix (Eq.~\ref{eq:rpa}) within the RPA.}
\label{fig:rpascheme}
\end{figure}

The interactions ($U$ and $J$ terms) are first written in magnetic channels: $-S_I \hat{\Gamma}(I;J) S_J$, where $I=(a\mu n)$ and $J=(b\nu m)$ with the orbital index ($a/b$), spin-direction index ($\mu/\nu$), and sublattice index ($n/m$). The spin operator is $S_{I=(a\mu n)}=\frac{1}{2} \sum_{\alpha\beta} a_{n\alpha}^\dag \sigma^\mu_{\alpha\beta} a_{n\beta}$, and the vertex function $\hat{\Gamma}(a\mu n;b\nu m)$ represented by Fig.~\ref{fig:rpascheme}(a) is diagonal in both spin-direction and sublattice subspaces and is given by
\begin{eqnarray}
\hat{\Gamma}(a\mu n;b\nu m)=\left\{
\begin{array}{ll}
2U \delta_{\mu\nu}\delta_{nm},&a=b \\
2J \delta_{\mu\nu}\delta_{nm},&a\neq b
\end{array} \right. .
\label{eq:vertex}
\end{eqnarray}

The RPA is a bubble summation as represented in Fig.~\ref{fig:rpascheme}(b). After solving the iterate equation we obtain the magnetic susceptibility matrix as
\begin{eqnarray}
\hat{\chi}(\bq,i\nu_n)=\left[\hat{
{\rm I}}-\hat{\Gamma}\hat{\chi}^{(0)}(\bq,i\nu_n)\right]^{-1}\hat{\chi}^{(0)}(\bq,i\nu_n).
\label{eq:rpa}
\end{eqnarray}
$\hat{\chi}^{(0)}(\bq,i\nu_n)$ is the bare susceptibility whose matrix element is defined as
\begin{eqnarray}
\hat{\chi}_{IJ}^{(0)}(\bq,i\nu_n)=\int_0^{1/T} \langle
S_{I}(-\bq,\tau)S_{J}(\bq,0) \rangle \mathrm{e}^{i\nu_n\tau} \mathrm{d}\tau .\nonumber \\
\end{eqnarray}
Here, we have used Matsubara frequency with $T$ the temperature.

\begin{figure}
\includegraphics[width=0.4\textwidth]{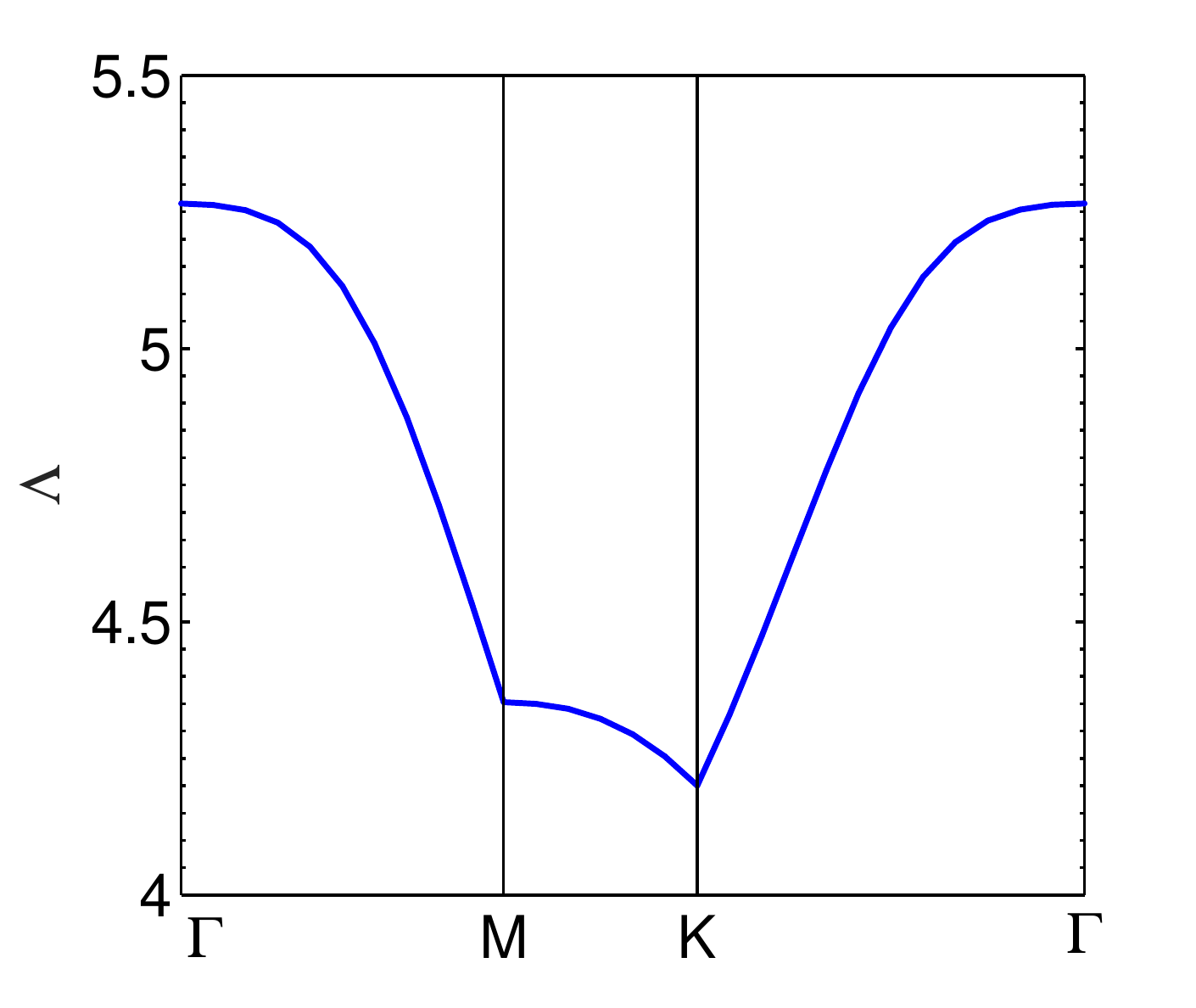}
\caption{The largest eigenvalue $\Lambda$ of $\hat{\chi}(\bq,\omega=0)$ along high symmetry cuts. Here $U=1.35$eV, $J=0.14$eV, $k_BT=0.3$eV.}
\label{fig:rpa}
\end{figure}

\begin{figure}
\includegraphics[width=0.3\textwidth]{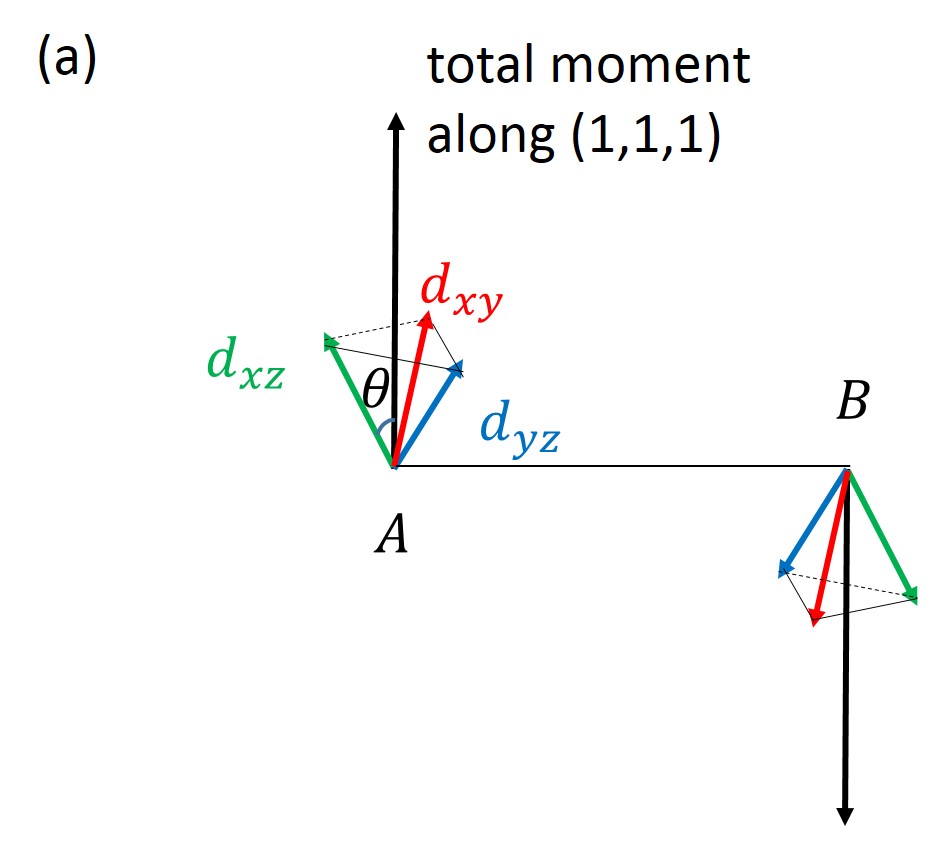}
\includegraphics[width=0.45\textwidth]{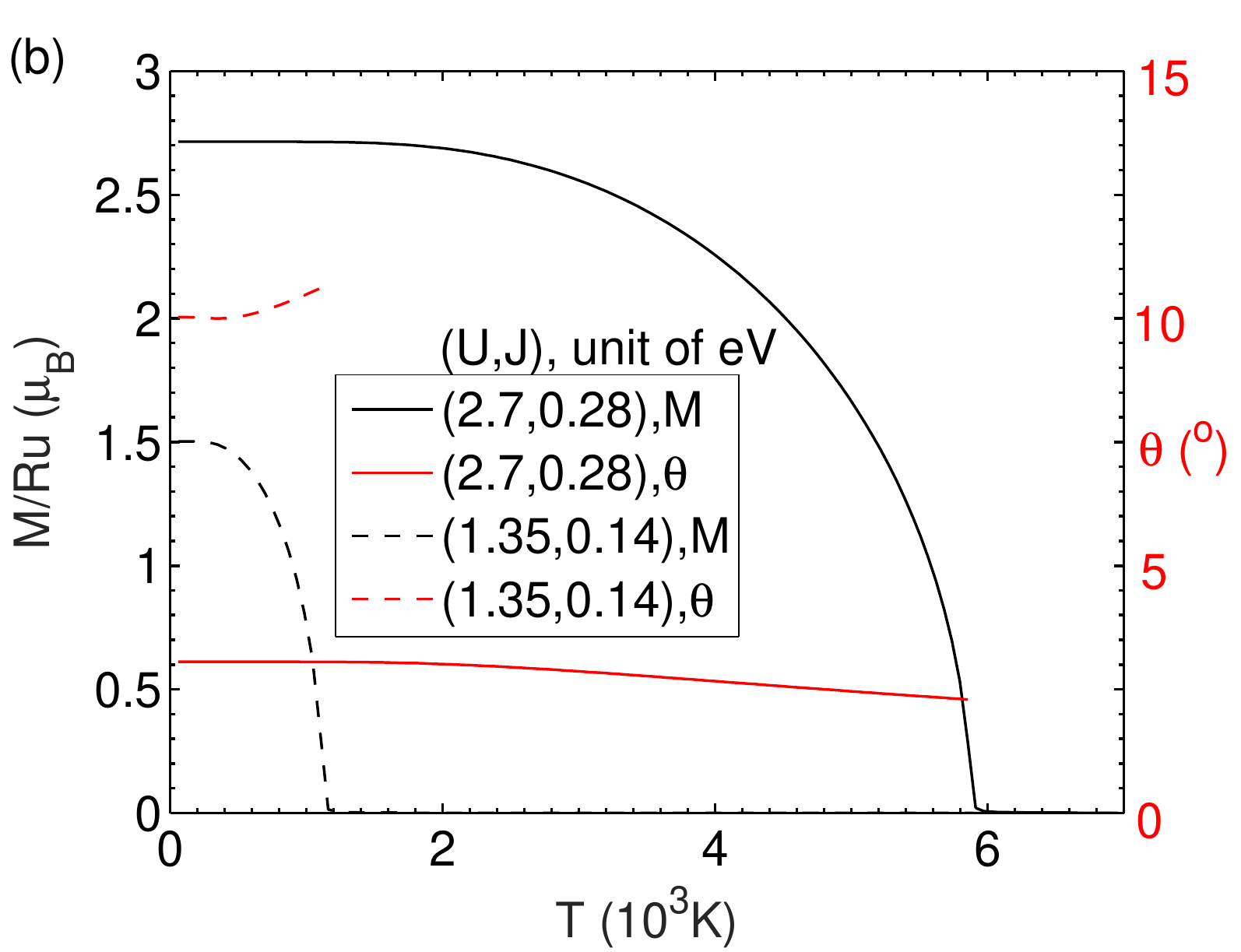}
\caption{(a)The AF configuration is schematically shown within a unit cell, with three orbital resolved moment represented by three different colors. The total moment is along c-axis [($1,1,1$)-direction]. (b)The total AF moment $M$ and the deviation angle $\theta$ vs temperature within the mean field approximation.}
\label{fig:meanfield}
\end{figure}

We use the interaction parameters $U=1.35$eV and $J=0.14$eV, being half of the values obtained by first principle calculation \cite{Tian2015}, since RPA is known to overestimate the magnetic instability.
We plot the leading (largest) eigenvalue $\Lambda$ of the hermitian susceptibility matrix $\hat{\chi}(\bq,0)$ as a function of $\bq$ in Fig.~\ref{fig:rpa}. The peak at $\bq=0$ implies a magnetic instability that is periodic across the unit cell. The corresponding leading eigenvector decides the form factor of the magnetic order, namely the magnetic structure within a unit cell. This is shown in Fig.~\ref{fig:meanfield}(a). We find the total moment (of the three orbitals) is along the $c$-axis and changes sign from one to the other sublattice within the unit cell. This is exactly the AF order seen in the neutron scattering experiment. More interestingly, the moments on the three orbitals are non-colinear about the $c$-axis. This is a prediction of the present work.

Next, we employ mean field theory to quantitatively investigate the AF moment and the transition temperature. The interactions are decoupled in magnetic channels:
\begin{eqnarray}
&&Un_{ia\uparrow}n_{ia\downarrow}\rightarrow -\frac{2U}{3}{\bf S}_{ia}\cdot {\bf S}_{ia} \nonumber \\
&&\rightarrow -\frac{4U}{3}\langle {\bf S}_{ia}\rangle \cdot {\bf S}_{ia} + \frac{2U}{3}\langle {\bf S}_{ia}\rangle\cdot \langle {\bf S}_{ia}\rangle ,
\end{eqnarray}
and
\begin{eqnarray}
&&J\sum_{\sigma\sigma'} a_{i\sigma}^\dag b_{i\sigma} b_{i\sigma'}^\dag a_{i\sigma'}\rightarrow  -2J{\bf S}_{ia}\cdot {\bf S}_{ib} \nonumber \\
&&\rightarrow -2J\langle{\bf S}_{ia}\rangle\cdot {\bf S}_{ib} -2J{\bf S}_{ia}\cdot \langle{\bf S}_{ib}\rangle +2J\langle{\bf S}_{ia}\rangle\cdot \langle{\bf S}_{ib}\rangle . \nonumber\\
\end{eqnarray}

Then we obtain the mean field Hamiltonian 
\begin{eqnarray}
H_{MF}&=&H_{0} - \frac{4U}{3}\sum_{ia}\langle {\bf S}_{ia} \rangle \cdot {\bf S}_{ia} \nonumber\\
&& - 2J\sum_{i,a\neq b}\langle {\bf S}_{ia} \rangle \cdot {\bf S}_{ib} ,
\end{eqnarray}
where $H_0$ is the non-interacting part, which is the same as Eq.~\ref{eq:Hk} but written in real space. The order parameters $\langle{\bf S}_{ia}\rangle=\frac{1}{2}\sum_{\alpha\beta}\langle a_{i\alpha}^\dag {\bf \sigma}_{\alpha\beta} a_{i\beta} \rangle$
are then numerically solved iteratively until convergence is achieved.

Our mean field result confirms the non-collinear AF configuration [Fig.~\ref{fig:meanfield}(a)] revealed by RPA in the normal state. We have performed {\it unrestricted} mean field calculations starting from random initial spin configurations. No translation symmetry is assumed in advance. However, the results all converge to the same non-collinear AF configuration up to a shift of the sublattice. The magnetic unit cell is always found to be equal to the lattice unit cell, which is in agreement with the unique peak $\bq=0$ in the leading eigenvalue of the RPA magnetic susceptibility in the momentum space as shown in Fig.~\ref{fig:rpa}. The non-collinear AF configuration is the result of the strong SOC on Ru atoms. The values of the total moment $M$ and the deviation angle $\theta$ of each orbital moment relative to c-axis [shown in Fig.~\ref{fig:meanfield}(a)] vs temperature are plotted in Fig.~\ref{fig:meanfield}(b). Two sets of parameters are used. One is $(U,J)=(2.7,0.28)$eV from first principle calculation \cite{Tian2015}, and the other is weakened by a factor of two, $(U,J)=(1.35,0.14)$eV. The latter leads to reduced $M$ and transition temperature. Since mean field theory overestimates the ordering, further considerations of quantum/thermal fluctuations beyond the mean field theory are necessary for quantitative comparison to experiments. Interestingly, we find $\theta$ becomes smaller with increasing interaction strength. This is because a larger $J$ tends to align the spins on different orbitals, while a larger SOC breaks the Hund's rule more significantly. Since neutron scattering only 'sees' the total moment, we expect this particular kind of orbital-resolved AF order can be observed in more delicate experiments like orbital-selective nuclear magnetic resonance\cite{Shimizu2012,Shimizu2015} through the anisotropic hyperfine interaction \cite{Kiyama2003} or orbital-resolved angle-resolved photoemission spectroscopy\cite{Cao2013} through the photon polarization selection rule \cite{Damascelli2003}.

\section{summary and future works}

In summary, we have constructed a $t_{2g}$-orbital model on a honeycomb lattice. Local Coulomb interaction was investigated in both RPA and mean field theory.
Experimentally observed Neel order is obtained.
Furthermore, our theory predicts that the magnetic moments on three orbitals are
non-collinear as a result of the strong spin-orbit coupling of Ru atoms. This particular kind of orbital-resolved AF order is expected to be observed in future experiments.

For future works, possible superconductivity in this compound after doping or under pressure is an interesting direction. Our $t_{2g}$-orbital model can be used as a minimal model to study possible superconductivity upon doping. The AF fluctuation may induce singlet Cooper pairing between the nearest neighbours. However, due to the strong SOC, triplet pairing may coexist with the singlet pairing. This part of work is being in progress.


\acknowledgements{The project was supported by NSFC (under grant No.10974086 and No.11023002) and the Ministry of Science and Technology of China (under grant No.2011CBA00108 and 2011CB922101). W. S. W. also thanks the support of the China Postdoctoral Science Foundation (under Grant No. 2014M561616). The numerical calculations were performed at the High Performance Computing Center of Nanjing University.}

\bibliography{Ru126}
\end{document}